# DIFFERENCE SEQUENCE COMPRESSION OF MULTIDIMENSIONAL DATABASES

István SZÉPKÚTI

ING Nationale-Nederlanden Hungary Insurance Co. Ltd.
H-1061 Budapest, Andrássy út 9, Hungary
e-mail: szepkuti@inf.u-szeged.hu

## Abstract

The multidimensional databases often use compression techniques in order to decrease the size of the database. This paper introduces a new method called difference sequence compression. Under some conditions, this new technique is able to create a smaller size multidimensional database than others like single count header compression, logical position compression or base-offset compression.

## 1. Introduction

### 1.1. Motivation

In this subsection, the relevance of multidimensional database compression is discussed. For the time being, it is enough to think of the multidimensional database as an $n$-dimensional matrix or array. Let $M$ denote an $n$-dimensional matrix and $M(i_1, i_2, \ldots, i_n)$ a cell of it ($i_1 \in D_1, i_2 \in D_2, \ldots, i_n \in D_n$). $D_j$ is the $j^{\text{th}}$ dimension of M, while the elements of $D_j$ are called dimension values ($j = 1, 2, \ldots, n$).

The total number of cells in a multidimensional database is the product of the number of dimension values for all dimensions $= \prod_{j=1}^{n} |D_j|$. On the other hand, these databases are often sparse, that is there are a lot of empty cells in them. Fortunately, there are several compression techniques which are able to remove a part or all of the empty cells from the multidimensional database.

Why should we compress multidimensional databases?

- The memory operations are significantly faster than hard disk operations, often by several orders of magnitude.
- The smaller the database, the more likely it will fit into the physical memory.
- In a virtual memory environment, the smaller the database, the more it will fit into the physical memory and the fewer virtual memory pagings will be necessary.

*1.2. Results*

The results of this paper can be summarized as follows:

- A new compression technique called difference sequence compression (DSC) is presented here.
- It is demonstrated that DSC results in a smaller database size than other compression techniques like single count header compression (SCHC), logical position compression (LPC) and base-offset compression (BOC), after making certain necessary assumptions.
- It is verified, using experiments on benchmark databases (TPC-D and APB-1), that DSC outperforms BOC.
- Even in the case of the APB-1 benchmark database, the multidimensional representation with DSC generally results in a smaller size than the corresponding table representation version compressed with different compression programs. There are only two exceptions: bzip2 and WinRAR.

*1.3. Related Work*

Several papers deal with the multidimensional model of databases and the compression of multidimensional databases. In this subsection, the most relevant ones are mentioned which are related to our subject in a way.

A thorough survey of logical multidimensional data models can be found in [15]. It does not concern itself with the possible compression of the multidimensional databases because it is more related to the physical representation than the logical model.

A variation of SCHC is described in [11]. In addition to this variation, the paper introduces two other compression techniques: LPC and BOC. With LPC, the size of the header can be decreased by 50%, when the size of SCHC header is maximal. BOC is able to decrease the header still further. In this paper, a new compression technique called DSC is introduced which is able to decrease the header even further in some circumstances. *Table 1* and *Table 2* are also from [11], except lines of the tables that show additional data on DSC and SCHC.

SCHC was introduced in [1], and this paper was also cited by several others, e.g. in [2, 3, 5, 8, 9, 13].

GOLDSTEIN et al. [2] introduce a page-level compression technique. Their basic observation is as follows. If we consider the actual range of values that appear in a given column on a given page, this is much smaller than the range of values in the underlying domain. If we know the range of potential values, we can represent any value in this range by storing just enough bits to distinguish between the values in this range.

The key ideas in the paper of GRAEFE et al. [3] are to compress attributes separately, to utilize the same compression scheme for each attribute of a domain, and to perform data manipulations before decompressing the data. In a simple performance comparison they showed that for data sets larger than the available memory, performance gains greater than the compression factor could be obtained as a larger fraction of the data could be retained in the workspace allocated to a query processing operator. With a compression factor of two, they obtained performance gains of a factor of three or better.

NG and RAVISHANKAR [5] present and then discuss three block-based database compression techniques. Two of them, BIT and ATS, are adaptations of conventional data compression techniques. The third one, TDC, exploits the redundancy among tuples in a different way to achieve compression. The idea of BIT is to compact every attribute value so that the entire tuple is bit-compressed (e.g. the values in the attribute $A = \{0, 1, \ldots, 899\}$ are represented in 10 bits instead of being represented by 3 ASCII characters). In the case of adaptive text substitution (ATS), if a sequence of symbols has occurred previously, it is replaced by a pointer to that previous occurrence. In [5] they employ ATS to compress and decompress a block of tuples.

In tuple differential coding (TDC), instead of storing tuples explicitly in tabular form as conventional databases do, we may capture and store the *differences* among them. If these differences require less space for storage on average than the original tuples, compression is achieved. Given a tuple $t$, it is converted to a unique integer $\varphi(t)$, which represents its *ordinal* position. Then all tuples are ordered by their $\varphi(t)$ values. The relation is partitioned into blocks (subsets). Within the blocks, using the first tuple as a reference, each successive tuple is replaced by its difference (in ordinals) with respect to its preceding one. In general, if $t_i$ and $t_j$ are consecutive tuples in a block, the latter tuple is replaced by $\varphi^{-1}(\varphi(t_j) - \varphi(t_i))$. Since the differences are numerically smaller than the tuples, they require fewer bytes of storage. The variable size differences are encoded by using *run-length encoding* to encode the number of leading zero components in each difference, thus achieving compression. The most important differences between TDC and DSC are listed below:

- TDC converts the entire tuple into an integer number ($\varphi(t)$ values), whereas with DSC only the primary key (logical positions) is converted.
- In TDC, the differences within a block are converted back to tuples, which are then compressed using run-length encoding. In DSC, differences are stored in the difference sequence without converting them back.
- The compressed difference tuples of TDC vary in size. Nevertheless, each element of the difference sequence occupies the same number of bits.

The study of RAY et al. [8] shows that attribute level compression is the best in a query processing perspective, but it has a poor compression ratio. They presented a modified attribute level compression algorithm based on non-adaptive arithmetic compression called COLA, which simultaneously provided good query processing and reasonable compression ratios. In the case of attribute level compression such as COLA, every attribute is compressed separately. DSC compresses each attribute in the primary key together to form a sequence of logical positions, which is then compressed still further.

In [9], SHOSHANI briefly mentions single count header compression in its original form as an example.

TOLANI and HARITSA [13] propose a new compression tool for XML documents called xGrind which directly supports queries in the compressed domain. A special feature of xGrind is that the compressed document retains the structure of the original document, permitting reuse of the standard XML techniques for processing the compressed document. xGrind simultaneously delivers improved query processing times and reasonable compression ratios.

The ITU-T Recommendation H.261 is specified in [4]. This video compression scheme uses two types of frames: $I$-frames (or intra-frames) and $P$-frames (or inter-frames). These frames might follow each other like the sequence:

$$IPPPIPPPP\ldots \quad (1)$$

The same sequence can be written as a regular expression: $(IP^*)^*$. The $P$-frames are pseudo-differences. The $I$-frames are related to the jumps of DSC, while the $P$-frames are associated with the positive elements of the difference sequence. In fact, DSC was inspired by the H.261 video compression method. The differences between H.261 and DSC are the following:

- H.261 is employed to compress video data, whereas in DSC a strictly increasing sequence of integer numbers is compressed.
- H.261 forms an alternating sequence of $I$-frames and $P$-frames. DSC uses a 'jump' sequence and a 'difference' sequence. A '0' in the difference sequence indicates that a jump is coming. For every $n^{th}$ jump, a pointer to the corresponding 0 element of the difference sequence is stored.

DSC may be viewed as a form of differential encoding, which is a kind of source encoding. (See [12] for example.) The source encoding is usually lossy, which is not a problem if, say, we want to compress a sound sample with differential encoding. However, there are a lot of applications of multidimensional databases where lossy compression is unacceptable. DSC was designed to be lossless. This is important because DSC compresses the header. We have to decompress the header without any losses in order to be able to determine the necessary physical position exactly.

Another multidimensional database compression method, called conjoint dimension, appeared in Express. Express was the first multidimensional analytical tool and dates back to 1970 [7]. Now it is a product of Oracle.

Let us suppose that the finite relation $R \subseteq D_1 \times \cdots \times D_n$ has a special property: some given elements of $D_1 \times \cdots \times D_h$ ($1 \leq h \leq k \leq n$ and the unique primary key of R is constituted by $D_1, \ldots, D_k$) cannot be found in the corresponding projection of R. So, in order to eliminate empty cells from the multidimensional array representation, we can define an equivalent $R'$ relation:

$$R' = \{((d_1, \ldots, d_h), d_{h+1}, \ldots, d_n) \mid ((d_1, \ldots, d_h), d_{h+1}, \ldots, d_n) \in$$
$$\in Conjoint \times D_{h+1} \times \cdots \times D_n \text{ such that } (d_1, \ldots, d_h, d_{h+1}, \ldots, d_n) \in R\}$$

where

$$Conjoint = \pi_{D_1,...,D_h}(R)$$

Here, $\pi$ denotes the projection operation of relations.

We have to be careful with conjoint dimensions. Consider, for example, the case when $h = k$, that is, all elements of the unique primary key are put into *Conjoint*. One can readily see that we could eliminate all empty cells this way and the multidimensional representation would become identical with the table-based one. (The multidimensional representation of a relation is a multidimensional array or matrix, whereas the table-based representation is nothing more than a table in a relational database.) Hence, we must exclude this extreme case of *Conjoint*, because it probably degrades the overall performance.

The paper of ZHAO et al. [16] introduced chunk-offset compression. In [16], the compressed multidimensional array occupied less space than the corresponding table representation of it. Moreover, at the same time, the compressed multidimensional array resulted in a faster operation than the table-based physical representation.

First, the n-dimensional array is divided into small n-dimensional chunks. Then the dense chunks (where the density $\rho > 40\%$) are stored without any modification. Sparse chunks are condensed using 'chunk-offset compression.' Here, just the existing data are stored using (*offsetInChunk*, *data*) pairs. This is the key idea behind the technique. In this compression method, not all the sparse cells are removed from the array. In the worst-case scenario when all chunks are just slightly denser than 40%, nearly 2.5 times more space is needed to store the cell values because all empty cells are also stored in this case. This may result in up to 2.5 times more disk input/output operation than absolutely necessary, when the chunks are read or written.

## 1.4. Organization

The rest of the paper is organized as follows. Section 2 describes three previously published compression techniques: single count header compression, logical position compression and base-offset compression. Section 3 introduces an improved method, that of difference sequence compression. The different compression techniques are compared with each other in Section 4. The theoretical results are then tested in experiments described in Section 5. Section 6 rounds off the discussion with a number of conclusions and suggestions for future study. Lastly, for completeness, we have the Acknowledgements, an appendix section and a references section.

## 2. Compression Techniques

Throughout this paper we employ the expressions 'multidimensional representation' and 'table representation,' which are defined as follows.

**Definition 1** Suppose we intend to represent relation $R$ physically. The multidimensional (physical) representation of $R$ is as follows:

- A compressed array, which only stores the nonempty cells, one nonempty cell corresponding to one element of $R$;
- The header, which is needed for the logical-to-physical position transformation;
- One array per dimension in order to store the dimension values.

The table (physical) representation consists of the following:

- A table, which stores all elements of relation $R$;
- A B-tree index in order to speed up the access to given rows of the table, when the entire primary key is given.

□

As it will be shown later, difference sequence compression (DSC) improves single count header compression (SCHC), logical position compression (LPC) and base-offset compression (BOC). These latter methods will be outlined in this section.

*Single count header compression.* By transforming the multidimensional array into a one-dimensional array, we obtain a sequence of empty and nonempty cells:

$$(E^*F^*)^* \qquad (2)$$

In the above regular expression, $E$ is an empty cell and $F$ is a nonempty one. The single count header compression (SCHC) stores only the nonempty cells and the cumulated run lengths of empty cells and nonempty cells. In [10], we used a variant of SCHC. The difference between the two methods is that the original method accumulates the number of empty cells and the number of nonempty cells separately. These accumulated values are placed in a single alternating sequence. The sum of two consecutive values corresponds to a logical position. (The logical position is the position of the cell in the multidimensional array before compression. The physical position is the position of the cell in the compressed array.) In [10], instead of storing a sequence of values, we chose to store pairs of a logical position and the number of empty cells up to this logical position: $(L_j, V_j)$. Just one pair is stored per $E^*F^*$ run, and $L_j$ points to the last element of the corresponding run. Let us suppose that we want to find the physical position $P(L)$ of logical position $L$ and $L_{j-1} < L \leq L_j$. Then the physical position

$$P(L) = \begin{cases} L - V_j, & \text{if } L_{j-1} + V_j - V_{j-1} < L; \\ \text{undefined}, & \text{otherwise.} \end{cases} \qquad (3)$$

$P(L)$ is undefined if and only if the cell at logical position $L$ is empty. Now let us assume that the physical position $P$ is given and $L_{j-1} - V_{j-1} < P \leq L_j - V_j$. We can compute its logical position $L(P)$ from the following simple formula:

$$L(P) = P + V_j. \tag{4}$$

In the rest of the paper when we mention SCHC we refer to the variant of this compression scheme defined in [10].

**Definition 2** The array storing the $(L_j, V_j)$ pairs of logical positions and number of empty cells will be called the SCHC header. □

The following two compression techniques improve SCHC when the SCHC header is maximal.

*Logical position compression.* The size of the SCHC header depends on the number of $E^*F^*$ runs. In the worst case there are $N = |R|$ runs, where $R$ is the relation which is represented multidimensionally using SCHC. Then the size of the SCHC header is $2N\iota$. (We assume that $L_j$ and $V_j$ are of the same data type and each of them occupies $\iota$ bytes of memory.) But in this case it is better to build another type of header. Instead of storing the $(L_j, V_j)$ pairs, it is more convenient to store just the $L_j$ sequence of each cell (that is not the $L_j$ sequence of runs).

The logical-to-physical position conversion part is carried out through a simple binary search. The physical position $P(L)$ of a logical position $L$ is defined as follows. The physical position

$$P(L) = \begin{cases} j, & \text{if there exists an } L_j \text{ such that } L = L_j; \\ \text{undefined}, & \text{otherwise.} \end{cases} \tag{5}$$

The physical-to-logical position conversion is just a simple lookup of an array element:

$$L(P) = L_P, \tag{6}$$

where $L(P)$ denotes the logical position of physical position $P$.

**Definition 3** The compression method which uses just the sequence of logical positions will be called logical position compression (LPC). The $L_j$ sequence used in logical position compression will be called the LPC header. □

The number of $E^*F^*$ runs is between 1 and $N = |R|$. Let $\nu$ denote the number of runs. Because the size of $L_j$ and $V_j$ is the same, the header is smaller with logical position compression if $\frac{N}{2} < \nu$. Otherwise, if $\frac{N}{2} \geq \nu$, the logical position compression does not result in a smaller header than the single count header compression. The header with logical position compression is half the size of SCHC header in the worst case, that is when $\nu = N$.

*Base-offset compression.* In order to store the entire $L_j$ sequence, we may need a huge (say 8-byte) integer number. However, the sequence is strictly increasing:

$$L_0 < L_1 < \cdots < L_{N-1}. \tag{7}$$

Here, $N$ denotes the number of elements in the $L_j$ sequence. The difference sequence, $\Delta L_j$, contains significantly smaller values. Based on this observation, we may compress the header still further.

Suppose that we need $\iota$ bytes to store one element of the $L_j$ sequence. In addition, there exists a natural number $l$ such that for all $k = 0, 1, 2, \ldots$ the

$$L_{(k+1)l-1} - L_{kl} \tag{8}$$

values may be stored in $\theta$ bytes and $\theta < \iota$. In this case we can store two sequences instead of $L_j$:

(i) $L_0, L_l, L_{2l}, L_{3l}, \ldots, L_{\lfloor \frac{N-1}{l} \rfloor l}$,

(ii) $L_0 - L_0, L_1 - L_0, \ldots, L_{l-1} - L_0$,
$L_l - L_l, L_{l+1} - L_l, \ldots, L_{2l-1} - L_l$,
$\ldots$,
$L_{\lfloor \frac{N-1}{l} \rfloor l} - L_{\lfloor \frac{N-1}{l} \rfloor l}, L_{\lfloor \frac{N-1}{l} \rfloor l+1} - L_{\lfloor \frac{N-1}{l} \rfloor l}, \ldots, L_{N-1} - L_{\lfloor \frac{N-1}{l} \rfloor l}$,

where $\lfloor x \rfloor$ means the integer part (floor) of $x$: $\lfloor x \rfloor = \max\{y \mid y \leq x \text{ and } y \text{ is integer}\}$.

**Definition 4** Sequence (i) will be called the base sequence, and sequence (ii) will be called the offset sequence. For convenience, let

$$B_k = L_{kl}, \tag{9}$$

$$O_j = L_j - B_{\lfloor \frac{j}{l} \rfloor}, \tag{10}$$

where $k = 0, \ldots, \lfloor \frac{N-1}{l} \rfloor$ and $j = 0, \ldots, N-1$. The compression method based on these two sequences will be labelled base-offset compression (BOC). The base and the offset sequences together will be called the BOC header. □

From the definition of the offset sequence, the following formula for the logical position follows immediately:

$$L_j = B_{\lfloor \frac{j}{l} \rfloor} + O_j. \tag{11}$$

Using this equation, we can define the logical-to-physical and the physical-to-logical position conversions as we did for LPC.

Now, let us compare the size of the LPC header to the BOC header. One element of the base sequence occupies $\iota$ bytes, while one offset sequence element requires $\theta$ bytes. The number of elements in the $L_j$ sequence is $N$. Thus, the space requirements of the two techniques are the following:

$$\text{LPC:} \quad N\iota,$$
$$\text{BOC:} \quad \left(\left\lfloor \frac{N-1}{l} \right\rfloor + 1\right)\iota + N\theta.$$

If N tends to $\infty$, one can clearly see that the multidimensional representation with BOC will occupy less space than that with LPC, if

$$\frac{\iota}{l} + \theta < \iota. \tag{12}$$

More details about these three compression techniques can be found in [1, 10, 11].

## 3. Improvements

The main new idea we propose here is that more flexibility is possible when an absolute address is stored, namely in case the relative address (offset) might be too large to store on given *s* bits.

The sequence of logical positions is strictly increasing:

$$L_0 < L_1 < \cdots < L_{N-1}.$$

In addition, the difference sequence $\Delta L_j$ contains smaller values than the original $L_j$ sequence. This property was utilized by the base-offset compression and will be used by the difference sequence compression, as well.

During the design of the data structures and the searching algorithm, the following principles were used:

- Compress the header so that the decompression is quick.
- It is not necessary to decompress the entire header.
- Searching can be done during decompression, and the decompression stops immediately when the header element is found or when it is demonstrated that the header element cannot be found (that is, when the corresponding cell is empty).

**Definition 5** Let us introduce the following notations.

$N$ is the number of elements in the sequence of logical positions ($N > 0$);
$L_j$ is the sequence of logical positions ($0 \leq j \leq N - 1$);
$\Delta L_0 = L_0$;
$\Delta L_j = L_j - L_{j-1}, j = 1, 2, \ldots, N - 1$;

The $D_i$ sequence ($D_i \in \{0, 1, \ldots, \overline{D}\}, i = 0, 1, \ldots, N - 1$) is defined as follows:

$$D_i = \begin{cases} \Delta L_i, & \text{if } \Delta L_i \leq \overline{D} \text{ and } i > 0; \\ 0, & \text{otherwise}; \end{cases} \tag{13}$$

where $\overline{D} = 2^s - 1$ and *s* is the size of one $D_i$ sequence element in bits.

The $J_k$ sequence will be defined recursively in the following way:

$$J_k = \begin{cases} L_0, & \text{if } k = 0; \\ L_j, & \text{otherwise, where } j = \min\{i \mid \Delta L_i > \overline{D} \text{ and } L_i > J_{k-1}\}. \end{cases} \quad (14)$$

Here the $D_i$ sequence is called the overflow difference sequence. There is an obvious deviation between $\Delta L_i$ and $D_i$, but the latter will also be called the difference sequence, if it is not too disturbing. As for $J_k$, it is called the jump sequence. The compression method using the $D_i$ and $J_k$ sequences will be called difference sequence compression (DSC). The $D_i$ and $J_k$ sequences together will be labelled as DSC header. □

**Definition 6** We are going to use the following: We say that $y$ is an $x$-bit unsigned integer, if $y \in \{0, 1, \ldots, 2^x - 1\}$. □

Using this notion we may assert that the elements of the $D_i$ sequence are $s$-bit unsigned integers.

Notice here that $\Delta L_i$ and $D_i$ are basically the same sequence. The only difference is that some elements of the original difference sequence $\Delta L_i$ are replaced with zeros, if and only if they cannot be stored in $s$ bits.

The difference sequence may also be called relative logical position sequence and we shall call the jump sequence the absolute logical position sequence.

From the definitions of $D_i$ and $J_k$, it can be seen that, for every zero element of the $D_i$ sequence, there is exactly one corresponding element in the $J_k$ sequence. For example, let us assume that $D_0 = D_3 = D_5 = 0$, and $D_1, D_2, D_4, D_6, D_7, D_8 > 0$. Then the aforementioned correspondence is shown in the following table:

| $D_0$ | $D_1$ | $D_2$ | $D_3$ | $D_4$ | $D_5$ | $D_6$ | $D_7$ | $D_8$ | ... |
|---|---|---|---|---|---|---|---|---|---|
| $J_0$ | | | $J_1$ | | $J_2$ | | | | ... |

From the previous definition, the recursive formula below follows for $L_j$.

$$L_j = \begin{cases} L_{j-1} + D_j, & \text{if } D_j > 0; \\ J_k, & \text{otherwise, where } k = \min\{i \mid J_i > L_{j-1}\}. \end{cases} \quad (15)$$

In other words, every element of the $L_j$ sequence can be calculated by adding zero or more consecutive elements of the $D_i$ sequence to the proper jump sequence element. For instance, in the above example

$$L_0 = J_0;$$
$$L_1 = J_0 + D_1;$$
$$L_2 = J_0 + D_1 + D_2;$$
$$L_3 = J_1;$$
$$L_4 = J_1 + D_4;$$

and so on.

Now we will show that there is a simple algorithm which is able to find $L$ in the $L_j$ sequence if the corresponding cell is not empty. In order to do this we need $A_k$ sequence of pointers which is defined as follows.

**Definition 7** For all $k$, $A_k = j$, if and only if $J_k = L_j$. We will refer to the $A_k$ sequence as the accelerator sequence. □

**Corollary 1** Suppose $J_k$ is an element of the jump sequence. Then the corresponding difference sequence element is $D_{A_k}$, which equals zero by definition. Thus, the accelerator sequence can be used to find the corresponding difference sequence element of a jump quite quickly.

```
int find_header(L)
{
  define and initialize variables;
  with binary search, find the jump element jump[k], for which
  L <= jump[k], and, if k-1 >= 0, jump[k-1] < L;
  if (L equals jump[k])
    return accelerator[k];
  else
  {
    if (k equals 0)
    {
      j = 0;
      decomp = jump[0];
    }
    else
    {
      j = accelerator[k-1];
      decomp = jump[k-1];
      while (decomp < L and j < size of the difference array - 1)
      {
        j = j + 1;
        decomp = decomp + difference[j];
      }
    }
    if (decomp equals L)
      return j;
    else
      return -1;
  }
}
```

*Fig. 1.*

The algorithm can be found in *Fig. 1*. In this algorithm, the previously defined sequences are denoted as follows: `jump[k]` = $J_k$, `difference[j]` = $D_j$,

accelerator[k] = $A_k$. During decompression, in the variable decomp, the sum of one jump sequence element and zero or more consecutive difference sequence elements is stored according to the recursive formula for $L_j$ given in (15). The algorithm returns a value of $j$, where $L = L_j$, if the cell at logical position $L$ is not empty. Otherwise, it returns a value of $-1$.

In order to save space, we can modify the above definition of $A_k$ and store only $A_0, A_n, A_{2n}, \ldots$, that is just every $n^{\text{th}}$ element of the original accelerator sequence. If we do so, the algorithm has to be altered slightly as well. For example, the else branch containing the while loop has to be modified like in *Fig. 2*.

```
...
else
{
  k = k - 1;
  k = k - k % n;         /* this ensures that k is a
                            multiple of n              */
  j = accelerator[k/n];  /* only every nth element of the
                            original accelerator sequence
                            is stored in this sequence  */
  decomp = jump[k];
  while (decomp < L and j < size of the difference array - 1)
  {
    j = j + 1;
    if (difference[j] equals 0)
    {
      k = k + 1;
      decomp = jump[k];
    }
    else
      decomp = decomp + difference[j];
  }
}
...
```

*Fig. 2.*

In this case, we have to expect zero difference sequence elements as well. When a zero comes, we will take the next element of the jump sequence. However, at the beginning of the algorithm it is enough to find $L$ with a binary search among the elements $J_0, J_n, J_{2n}, \ldots$ because the accelerator sequence only contains pointers for these jumps.

The accelerator sequence is a useful method for speeding up the retrieval (point query) operation for the following reasons:

- It is not necessary to store the accelerator sequence on the hard disk since it can be populated easily based on the difference sequence in one pass. This is needed only once, after the difference array is loaded from the hard disk into the memory.

- In practice it does not increase the memory requirements significantly, as it can be seen from the argument below.

Now, we are going to compare the size of the accelerator sequence to the size of the jump sequence. Let us suppose that the size of one jump sequence element is $\iota$, where one accelerator sequence element is $\alpha$. The jump sequence has $M$ elements (i.e. there are $M$ jumps). In addition, let us assume that only every $n^{\text{th}}$ element of the accelerator sequence is stored, that is $A_0, A_n, A_{2n}, \ldots$

$$\frac{\text{size of the accelerator sequence}}{\text{size of the jump sequence}} = \frac{(\lfloor \frac{M-1}{n} \rfloor + 1)\alpha}{M\iota} \leq \frac{(\frac{M-1}{n} + 1)\alpha}{M\iota} = \quad (16)$$

$$= \left(\frac{1}{n} + \frac{1}{M} - \frac{1}{Mn}\right)\frac{\alpha}{\iota} \to \frac{\alpha}{n\iota}, \text{ if } M \to \infty. \quad (17)$$

In the experiments performed the parameters had the following values: $\alpha = 32$ and $n = 16$. The third variable, $\iota$, was 64 for the TPC-D benchmark database, and 32 for the APB-1 database. Hence the value of $\frac{\alpha}{n\iota}$ was 3.1% or 6.3%. But the jump sequence is just one part of the multidimensional representation, so the accelerator sequence did not increase the memory requirements significantly.

## 4. Comparison of Compression Techniques

The difference between BOC and DSC is that

- the logical position $L_j$ is calculated as the sum of a base array element and an offset array element in BOC, while
- $L_j$ is derived from the jump and the difference arrays for DSC.

It is easy to see that the number of elements in the offset array and the difference array is just the same, but are there fewer jumps than base array elements? The answer to this question is that there are not more jumps than base array elements in the case when the size of one offset array element ($\theta$) is equal to or less than the size of one difference array element ($\zeta$) or, written in symbols,

$$\theta \leq \zeta.$$

Throughout this section the sizes will be measured in bits. We assume that we insert a jump only when it is absolutely necessary, that is when the difference between two consecutive elements of the $L_j$ sequence of logical positions is so large that it cannot be stored in $\zeta$ bits. This assumption is really a direct consequence of the definition of the jump sequence.

**Theorem 1** *There are never more jumps than base array elements if $\theta \leq \zeta$.*

*Proof.* Assume that there are $b$ elements in the base array, where $b = \lfloor \frac{N-1}{l} \rfloor + 1$. The base array elements partition the $L_j$ sequence of logical positions into $b$ buckets. For example, let us take the following sequences:

| $B_0$ | | $B_1$ | | | | $B_2$ | ... | $B_{b-1}$ | |
|---|---|---|---|---|---|---|---|---|---|
| $L_0$ | $L_1$ | $L_2$ | $L_3$ | $L_4$ | $L_5$ | $L_6$ | ... | $L_{N-2}$ | $L_{N-1}$ |

With the above sequences, the buckets are defined as follows:

$$\begin{aligned} bucket_0 &= \{L_0, L_1, L_2\}, \\ bucket_1 &= \{L_3, L_4, L_5\}, \\ &\ldots \\ bucket_{b-1} &= \{L_{N-2}, L_{N-1}\}. \end{aligned}$$

Let us suppose indirectly that there are more jumps than base array elements ($b$). It means that there is at least one bucket that has at least two jumps. Let us assume that $bucket_i$ contains two jumps at $L_j$ and $L_k$ ($L_j < L_k$). The following inequality obviously holds:

$$B_i \leq L_j < L_k. \tag{18}$$

In addition to this, if $i < b - 1$, then

$$L_k < B_{i+1}. \tag{19}$$

$B_i$, $L_j$ and $L_k$ are in the same bucket: $B_i, L_j, L_k \in bucket_i$. It means that the $L_k - B_i$ value can be stored in $\theta$ bits. It also means that one $\zeta$-bit unsigned integer is sufficient to store all of these elements ($\zeta \geq \theta$):

$$L_{j+1} - L_j, L_{j+2} - L_{j+1}, \ldots, L_{k-1} - L_{k-2}, L_k - L_{k-1}. \tag{20}$$

That is at $L_k$, there must not be a jump, which is a contradiction. This implies that there are never more jumps than base array elements if $\theta \leq \zeta$. ∎

**Corollary 2** The multidimensional representation with DSC does not result in a bigger database size than with BOC if $\theta = \zeta$. (The size of one offset array element is just the same as the size of one difference array element.)

If we decrease the size of one difference array element from $\zeta_1$ to $\zeta_2$ ($\zeta_1 > \zeta_2 > 0$), then the number of jumps will increase in such a way that new jumps will be inserted between old ones. How many new jumps will be inserted? We have to insert a new jump for all difference sequence elements that do not fit into a $\zeta_2$-bit unsigned integer.

On the one hand, the number of jumps will increase from $M_1$ to $M_2$ ($M_1 < M_2$), but on the other, the size of one difference sequence element will decrease from $\zeta_1$ to $\zeta_2$.

**Definition 8** Let us introduce the distribution function of $\Delta L_j$ as follows:

$$F(x) = P(\Delta L_j < x). \tag{21}$$

□

**Theorem 2** *Assume that we store the difference array elements in $\zeta_i$-bit unsigned integers ($i = 1, 2$) and that we decrease the size of the difference sequence elements from $\zeta_1$ to $\zeta_2$ ($\zeta_1 > \zeta_2 > 0$). The jump sequence elements are stored in $\iota$ bits ($\iota > 0$). If the slope of the $F(2^x)$ function between $\zeta_2$ and $\zeta_1$ is less than $\frac{1}{\iota}$, then we are going to obtain a smaller multidimensional representation after the decrease of the difference array.*

*Proof.* We store the difference array elements as $\zeta_i$-bit unsigned integers ($i = 1, 2$). This means that the number of jumps is

$$M_i = P(\Delta L_j \geq 2^{\zeta_i})N = (1 - F(2^{\zeta_i}))N. \tag{22}$$

How many new jumps will be inserted if we decrease the size of one difference sequence element from $\zeta_1$ to $\zeta_2$? The answer is

$$M_2 - M_1 = (1 - F(2^{\zeta_2}))N - (1 - F(2^{\zeta_1}))N = (F(2^{\zeta_1}) - F(2^{\zeta_2}))N. \tag{23}$$

The cost of changing the size of one difference sequence element is $(M_2 - M_1)\iota$ bits, because the size of the jump array will increase with this. On the other hand, the benefit of this modification is $N(\zeta_1 - \zeta_2)$. The latter formula stands for the size difference of the old and the new difference sequence. Let us examine when the benefit is larger than the cost:

$$N(\zeta_1 - \zeta_2) > (M_2 - M_1)\iota \tag{24}$$

$$N(\zeta_1 - \zeta_2) > (F(2^{\zeta_1}) - F(2^{\zeta_2}))N\iota \tag{25}$$

$$\frac{\zeta_1}{\iota} - \frac{\zeta_2}{\iota} > F(2^{\zeta_1}) - F(2^{\zeta_2}) \tag{26}$$

$$\frac{1}{\iota} > \frac{F(2^{\zeta_1}) - F(2^{\zeta_2})}{\zeta_1 - \zeta_2} \tag{27}$$

That is, the benefit is bigger than the cost when the slope of the $F(2^x)$ function between $\zeta_2$ and $\zeta_1$ is less than $\frac{1}{\iota}$. ∎

*Remark.* The sizes were measured in bits. For efficiency reasons we may not want to allow all the possible sizes. For example, we might restrict the domain to 8, 16, 32 and 64. This is exactly what we did in our experiments because the programming language had built-in support for 8-, 16-, 32- and 64-bit unsigned integers.

## 5. Experiments

We carried out experiments in order to check whether DSC is able to create a smaller-sized database than the previously published BOC. We also tested how quickly we could retrieve the contents of cells when the database was compressed using DSC. The hardware and software components used for our experiments are listed in **Appendix**.

In the experiments we made use of two benchmark databases: TPC-D [14] and APB-1 [6]. One relation was derived per benchmark database in exactly the same way as described in [11]. Then these relations were stored physically with a multidimensional representation and table representation.

When we compare the difference sequence compression of the multidimensional representation of relation $R$ to compressions of the table representation of relation $R$, we obtain an interesting result. (Here $R$ is a relation derived from one of the benchmark databases: TPC-D or APB-1.) Both *Table 1* and *Table 2* show the difference sequence compression results in a smaller multidimensional representation than base-offset compression. In the case of the TPC-D benchmark database, the multidimensional representation with BOC turned out to be already smaller than all those for alternative compression techniques of the table representation (see [11]).

In the APB-1 benchmark database, BOC was less successful. It produced a slightly bigger database than the compressions of the table representation. However, with the exception of bzip2 and WinRAR, DSC outperformed the other compressors.

*Table 1*. TPC-D benchmark database

| **Compression** | **Size in bytes** | **Percentage** |
|---|---|---|
| **Table representation** | | |
| Uncompressed | 279,636,324 | 100.0% |
| ARJ | 92,429,088 | 33.1% |
| gzip | 90,521,974 | 32.4% |
| WinZip | 90,262,164 | 32.3% |
| PKZIP | 90,155,633 | 32.2% |
| jar | 90,151,623 | 32.2% |
| bzip2 | 86,615,993 | 31.0% |
| WinRAR | 81,886,285 | 29.3% |
| **Multidimensional representation** | | |
| Single count header compression | 145,256,792 | 51.9% |
| Base-offset compression | 74,001,692 | 26.5% |
| Difference sequence compression | 67,925,100 | 24.3% |

It is worth making a comparison here with SCHC as well. In TPC-D, SCHC

*Table 2*. APB-1 benchmark database

| Compression | Size in bytes | Percentage |
|---|---|---|
| **Table representation** | | |
| Uncompressed | 1,295,228,960 | 100.0% |
| jar | 124,462,168 | 9.6% |
| gzip | 124,279,283 | 9.6% |
| WinZip | 118,425,945 | 9.1% |
| PKZIP | 117,571,688 | 9.1% |
| ARJ | 115,085,660 | 8.9% |
| bzip2 | 99,575,906 | 7.7% |
| WinRAR | 98,489,368 | 7.6% |
| **Multidimensional representation** | | |
| Base-offset compression | 125,572,184 | 9.7% |
| Difference sequence compression | 113,867,897 | 8.8% |
| Single count header compression | 104,959,936 | 8.1% |

resulted in an almost twofold database size than BOC. Whereas with the APB-1 benchmark database, SCHC beat even DSC (but the multidimensional representation with SCHC is still bigger than the table representation compressed with bzip2 or WinRAR). The reason for this is that almost every run in the TPC-D database contains just one nonempty cell. So the number of runs was almost maximal. Whereas the tested relation in APB-1 is a time series, the runs containing at least 17 consecutive nonempty items. This kind of relation can be compressed much better with SCHC than the other available methods.

The speed of retrieval was also tested in the two benchmark databases just like in [11]. Random samples were taken from the tested relations. The sample sizes were 100, 500, 1,000, 5,000, 10,000, 50,000 and 100,000 in the experiments. Then, with point queries, we searched for these sample elements one by one in the table representation with the help of the B-tree index, and in the multidimensional representation using the DSC header. The elapsed time was measured in seconds. *Table 3* shows the results of the experiments in the case of DSC.

The first column shows the sample size. Columns 2 – 4 relate to the TPC-D benchmark database, columns 5 – 7 to APB-1. Columns 2 and 5 tell us how long it took to find the sample elements in the table representation. Columns 3 and 6 do the same for the multidimensional representation with DSC. Columns 4 and 7 are quotients. They demonstrate how many times faster the multidimensional representation was than the table representation in our experiments. These quotients, as functions of the sample size, are also depicted in *Fig. 3*.

As it can be seen in *Table 3* and *Fig. 3*, the multidimensional representation with DSC was 1.3 – 18.8 times faster than the table representation. The reason behind this is that, in an ideal situation, the entire DSC header can be loaded into the physical memory. Then we have to read from the hard disk only once in order to retrieve the content of the cell if we want to execute a point query in the multidimensional representation. With the table representation, first we have to find the record ID using the B-tree index. Afterwards we can retrieve the necessary record. This process may need significantly more hard disk operations than the previous one.

*Table 3.* The speed of retrieval in the case of DSC

| | **TPC-D** | | | **APB-1** | | |
|---:|---:|---:|---:|---:|---:|---:|
| **Sample** | **Table** | **Array** | **Quotient** | **Table** | **Array** | **Quotient** |
| 100 | 3.0 | 1.4 | 2.1 | 4.6 | 1.8 | 2.5 |
| 500 | 9.9 | 6.3 | 1.6 | 13.0 | 7.0 | 1.9 |
| 1,000 | 15.4 | 11.6 | 1.3 | 23.1 | 13.6 | 1.7 |
| 5,000 | 123.0 | 48.8 | 2.5 | 189.5 | 63.4 | 3.0 |
| 10,000 | 308.8 | 87.0 | 3.5 | 420.4 | 115.5 | 3.6 |
| 50,000 | 1,489.6 | 150.5 | 9.9 | 2,115.9 | 416.2 | 5.1 |
| 100,000 | 2,891.8 | 154.1 | 18.8 | 4,231.1 | 794.7 | 5.3 |

*Table 4.* The speed of retrieval in the case of SCHC

| | **TPC-D** | | | **APB-1** | | |
|---:|---:|---:|---:|---:|---:|---:|
| **Sample** | **Table** | **Array** | **Quotient** | **Table** | **Array** | **Quotient** |
| 100 | 4.7 | 6.2 | 0.8 | 5.3 | 1.9 | 2.8 |
| 500 | 23.9 | 25.4 | 0.9 | 15.3 | 7.3 | 2.1 |
| 1,000 | 40.7 | 44.0 | 0.9 | 26.1 | 14.2 | 1.8 |
| 5,000 | 173.4 | 161.4 | 1.1 | 223.0 | 62.2 | 3.6 |
| 10,000 | 329.3 | 311.9 | 1.1 | 442.6 | 106.0 | 4.2 |
| 50,000 | 1580.8 | 1644.5 | 1.0 | 2173.9 | 331.1 | 6.6 |
| 100,000 | 3196.5 | 4029.7 | 0.8 | 4344.5 | 620.7 | 7.0 |

*Table 4* and *Fig. 4* show the results for SCHC in a manner similar to that of DSC. Curiously, SCHC was unable to speed up the retrieval operation in the TPC-D database. In APB-1, the speed-up factor was between 1.8 and 7.0. This behaviour can be accounted for by the size differences between the two compressions: DSC and SCHC.

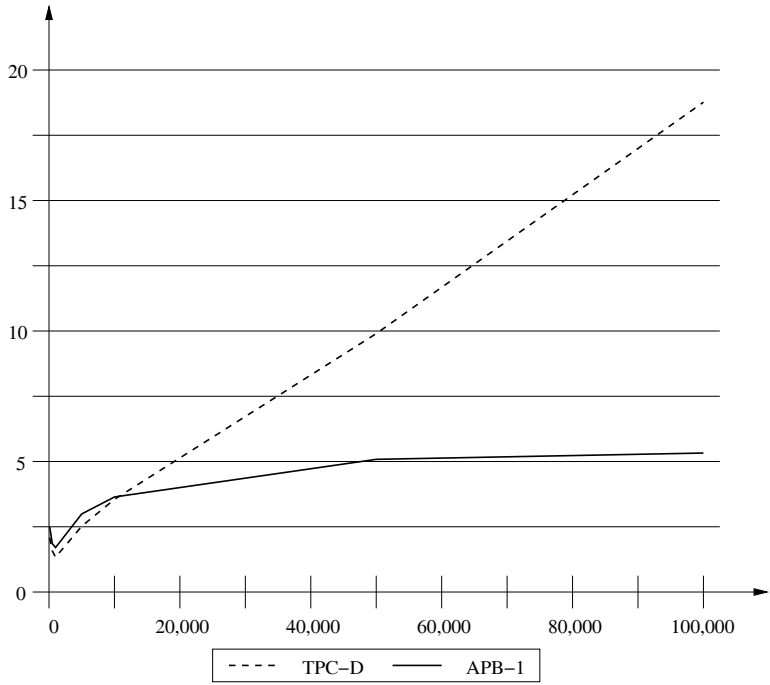

*Fig. 3.* The speed of retrieval for DSC as a function of the sample size

## 6. Conclusion

In this paper we introduced a new compression method called difference sequence compression (DSC).

It has been shown that the difference sequence compression is able to create a smaller multidimensional physical representation than other previously published methods such as single count header compression (SCHC), logical position compression (LPC) and base-offset compression (BOC) under certain conditions. Difference sequence compression does not result in a bigger multidimensional representation than base-offset compression if the size of the difference sequence is exactly the same as the size of the offset sequence. Using the distribution function of the original difference sequence ($\Delta L_j$), a condition was given when it is beneficial to change the size of the difference sequence elements to save space.

From experiments on benchmark databases (TPC-D and APB-1) we were able to verify that difference sequence compression can beat base-offset compression, producing improvements of some 8–9% of the multidimensional database compressed with BOC.

Even with the APB-1 benchmark database, the multidimensional representation with difference sequence compression resulted in a smaller size than the table representation compressed with different compression programs. There were only two exceptions: bzip2 and WinRAR.

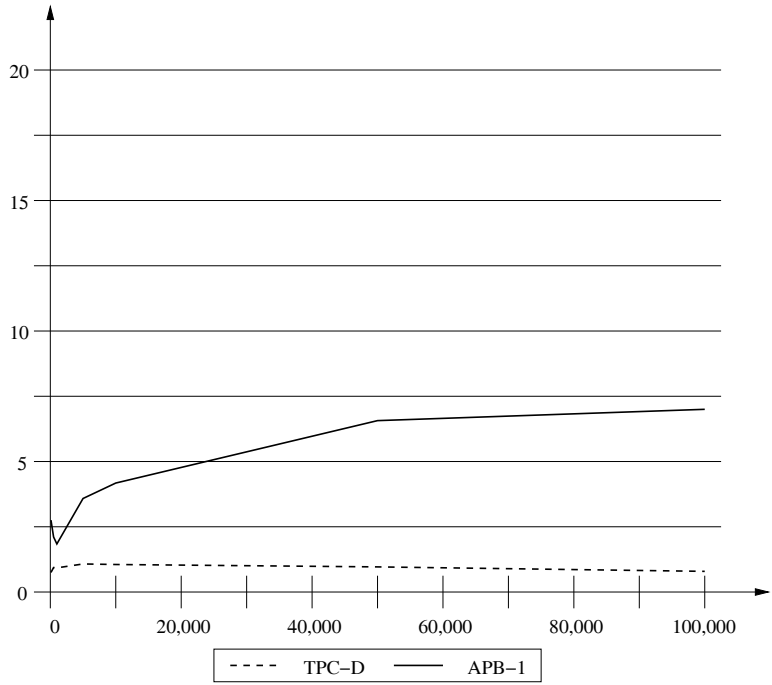

*Fig. 4.* The speed of retrieval for SCHC as a function of the sample size

The decompression algorithm of difference sequence compression seems more complicated than the similar algorithm of base-offset compression. This might cause performance degradation because of the increase in the number of operations. In fact, the base-offset compression resulted in a 1.4 – 7.8 times faster operation than the table representation, as shown in [11]. On the other hand, the difference sequence compression was 1.3 – 18.8 times quicker than the table physical representation. Thus, the new method still performs much better than the table-based solution.

DSC was compared with SCHC as well. In the case of the APB-1 benchmark database, SCHC was more successful than DSC, because it resulted in a somewhat smaller compressed multidimensional representation and faster operation. However, SCHC was much worse for TPC-D. We may infer from this that SCHC is much more sensitive to the composition of runs (whether the runs contain a lot of consecutive nonempty elements or not). This sensitivity does not exist in the case of LPC, BOC or DSC since they store the logical positions of all nonempty cells and not just one logical position per $E^*F^*$ run.

In several applications of multidimensional databases speed is extremely im-portant. Consider, for example, a decision support system that provides dynamic reports and on-line analytical capabilities to its users. This system has to answer the queries within seconds or a split second. Otherwise the users will become impatient and lose their interest in the system. And a system without users is not of much worth.

With multidimensional databases just like in many other databases, one important goal is to minimize the number of input/output operations in order to speed up the application. In an ideal situation, a compressed multidimensional physical representation is able to take the number of input/output operations close to the absolutely necessary minimum. In order to achieve this, we have to find better and better compression methods, which can make smaller and smaller databases, while maintaining their speed advantage compared to the table physical representation of databases.

## Acknowledgements

I would like to thank Prof. Dr. János Csirik for his invaluable comments on earlier versions of this paper and his very useful suggestions.

## Appendix

The table below shows hardware and software used for testing. The speed of the processor, the memory and the hard disk influences the experimental results significantly, as so does the size of the memory. In the computer industry, all of these parameters increase quickly over time. Just the rise of hard disk speed is somewhat slower. Hence, it is expected that the presented results remain valid despite the continuing improvement of the computers.

| | |
|---|---|
| Computer | Toshiba Satellite 300CDS |
| Processor | Intel Pentium MMX |
| Processor speed | 166 MHz |
| Memory size | 80 MB |
| Hard disk manufacturer | IBM |
| Hard disk size | 11 GB |
| File system | ext2 |
| Page size of B-tree | 4 KB |
| Operating system | Red Hat Linux release 6.2 (Zoot) |
| Kernel version | 2.2.14-5.0 |
| Compiler | gcc version egcs-2.91.66 19990314/Linux |
| Programming language | C |